# Topology Optimization in Three-Phase C-Core SRMs

Gholamreza Davarpanah
*Department of Electrical Engineering*
Amirkabir University of Technology
Tehran, Iran
ghr.davarpanah@aut.ac.ir

Sajjad Mohammadi
*Department of Electrical Engineering and Computer Science*
MIT
Cambridge, USA
sajjadm@mit.edu

James L. Kirtley
*Department of Electrical Engineering and Computer Science*
MIT
Cambridge, USA
kirtley@mit.edu

*Abstract*—This paper investigates the optimized combination of rotor and stator teeth in a three-phase switched reluctance motor featuring a connected C-core topology to attain a larger winding area and, thus, a higher electrical loading capability, leading to a higher torque density. The general formulation is discussed. Additionally, the shorter flux path within the design leads to a reduced core loss. The finite element method is employed in the design and field analysis. A comparison with a conventional motor shows the superiority of the proposed configuration. The optimized design is prototyped and tested. Both static and dynamic torques are extracted. A good correlation between simulation and experimental results is observed.

TABLE I. DIMENSIONS OF STUDIED SRMs

| Parameter | SRMs | | |
|---|---|---|---|
| | 12/10 | 12/14 | 12/16 |
| Stator outer diameter, $D_o$ (mm) | 82 | 82 | 82 |
| Stator yoke thickness, $b_{sy}$ (mm) | 3.57 | 2.55 | 2.23 |
| Stator pole height, $h_s$ (mm) | 10.93 | 11.95 | 12.27 |
| Stator pole arc, $\beta_s$ (deg) | 14.14 | 10.1 | 8.83 |
| Rotor pole height, $h_r$ (mm) | 3.06 | 3.65 | 4.17 |
| Rotor shaft diameter, $D_{sh}$ (mm) | 14 | 14 | 14 |
| Rotor pole arc, $\beta_r$ (deg) | 13.44 | 9.60 | 8.40 |
| Air-gap length, $l_g$ (mm) | 0.17 | 0.17 | 0.17 |
| Stack length, $L$ (mm) | 25.4 | 25.4 | 25.4 |
| Number of turns per pole, $T_{pole}$ | 60 | 90 | 80 |
| Steel type | M19-24G (knee point = 1.9T) | | |

## I. INTRODUCTION

Switched reluctance motors (SRMs) are appealing for their robust structure, low maintenance costs, and affordability [1]-[9]. They have applications in electric and hybrid electric vehicles, home appliances, MIR machines, surgical tools, aircraft actuation systems, etc. There have been attempts to propose new designs with increased torque density and reduced torque ripple [10]-[15]. One recent approach has been embedding permanent magnets into the SRM structure [16]-[18]. Another technique is to optimize the combination of rotor and stator teeth [19]-[25]. A two-phase E-core SRM was introduced in [26], showing a significant increase in mean torque, which is extended to another design with a better structural strength using connected E-cores [27]-[28]. A new equation for determining the number of stator and rotor teeth was introduced in [29], resulting in a configuration with more rotor teeth than stator teeth. A conventional six-phase SRM that can be excited as a three-phase or a six-phase motor is proposed in [30]. Other topologies for C-core SRMs, performing better than the conventional ones, are offered in [31]-[32]. In [33], a technique is proposed to reduce torque ripple.

This paper investigates the optimized combination and the general formulation of rotor and stator teeth in a three-phase SRM with a connected C-core configuration proposed in [34] to maximize the winding area to achieve a higher electrical loading capability and, thus, a higher torque density. Additionally, the main flux path within the C-core design is shorter, leading to a reduction in core losses. Three motors with different combinations of rotor and stator teeth are investigated and compared, where each motor is separately optimized. Also, FEM is employed in the design and field analyses. A comparison with the conventional motor is also made. Finally, the selected SRM is prototyped. The experimental results validate the design as well as the simulation results for both static and dynamic tests.

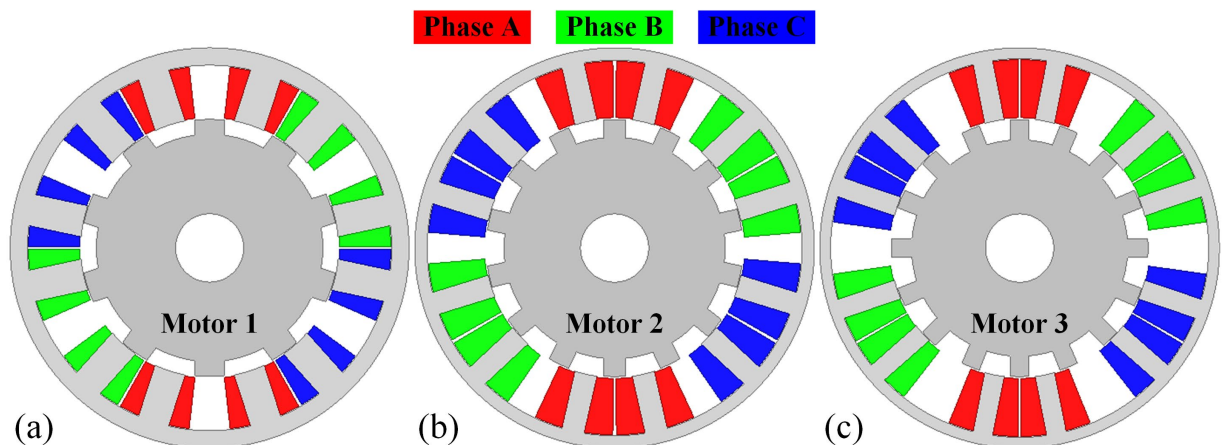

Fig. 1. Studied C-core SRMs: (a) 12/10, (b) 12/14, and (c) 12/16.

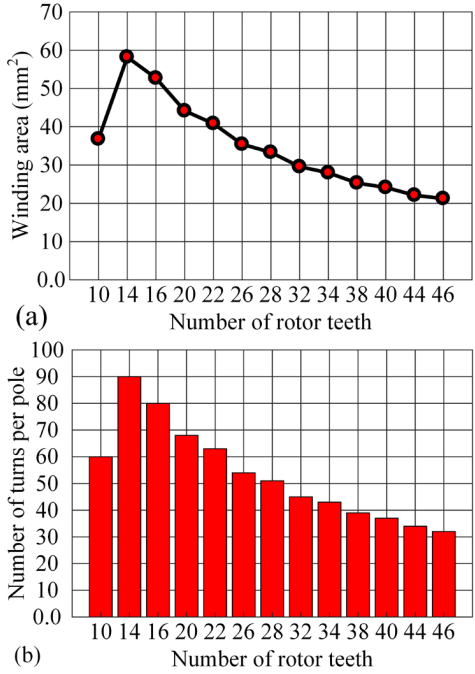

Fig. 2. Sensitivity of (a) winding area and (b) stator turns to rotor teeth number

## II. TOPOLOGY OF THE MOTORS

As shown in Fig. 1, three C-core SRMs with different combinations of rotor and stator teeth are investigated, in which each phase contains two C-cores, which are placed on the opposite sides of the rotor ($m$=2) to maintain symmetry and thus cancel out radial forces on the rotor. The specifications of the studied SRMs are summarized in Table I. The number of stator teeth obtained is as follows:

$$N_s = 2mq \quad (1)$$

where $q$ is the number of phases, $m$ is the number of C-cores per phase, and the factor of 2 is due to the C-core having two teeth. The number of rotor teeth is obtained as shown below:

$$N_r = 2m + 2n; \; n = 2, 3, 5, 6, 8, 9, 11, 12, ... \quad (2)$$

It should be noted that no value of $n$ works. For example, in a three-phase SRM ($q$=3) having opposite C-core of the same phase ($m$=2), the minimum $N_r$ should be 8 because otherwise, stator teeth of adjacent phases would intersect with each other, so $n$ cannot be 1. Generally, any value of $n$ leading to a value of $N_r$, which is a multiple of 6, doesn't work because all rotor teeth would be aligned with all stator teeth, and thus, no reluctance torque can be generated. For example, $n$=4 results in a 12/12 SRM. To have a fair comparison, at a fixed motor volume, i.e., the same outer diameter Do, and stack length L, the parameters of each motor are optimized to attain the highest torque. As shown in Fig. 2, it can be seen that the 12/14 SRM offers more winding space and, thus, a higher number of turns compared to the 12/10 and 12/16 SRMs.

TABLE II. MEAN AND PEAK TORQUES

| | | Phase current (A) | | | | |
|---|---|---|---|---|---|---|
| | | 1 | 2 | 3 | 4 | 5 |
| Mean torque (N.m) | 12/10 SRM | 0.025 | 0.100 | 0.221 | 0.370 | 0.528 |
| | 12/14 SRM | 0.051 | 0.195 | 0.389 | 0.597 | 0.807 |
| | 12/16 SRM | 0.043 | 0.170 | 0.348 | 0.533 | 0.712 |
| Peak torque (N.m) | 12/10 SRM | 0.034 | 0.137 | 0.308 | 0.532 | 0.785 |
| | 12/14 SRM | 0.070 | 0.278 | 0.597 | 0.954 | 1.297 |
| | 12/16 SRM | 0.062 | 0.245 | 0.536 | 0.864 | 1.181 |

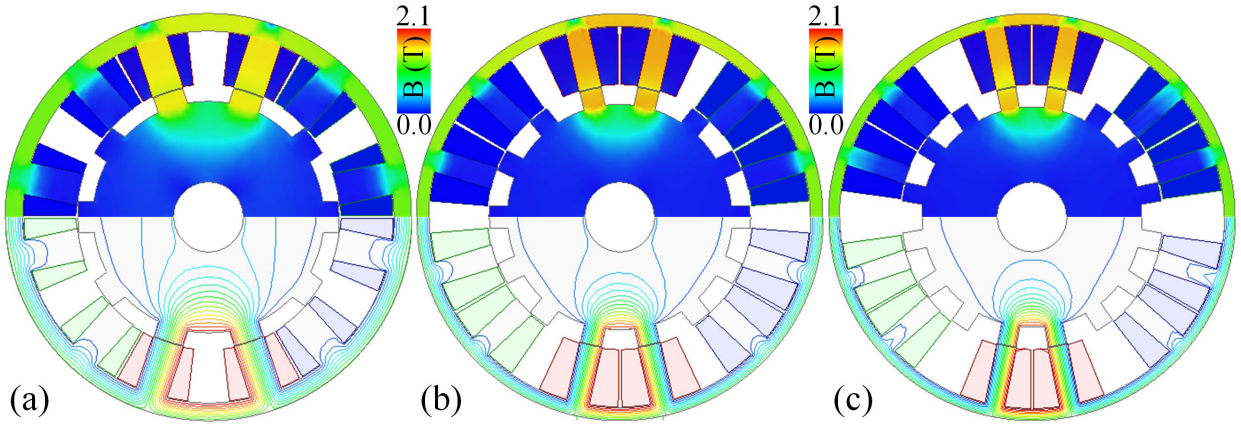

Fig. 3. Magnetic flux density distributions and flux lines: (a) 12/10 SRM, (b) 12/14 SRM, and (c) 12/16 SRM.

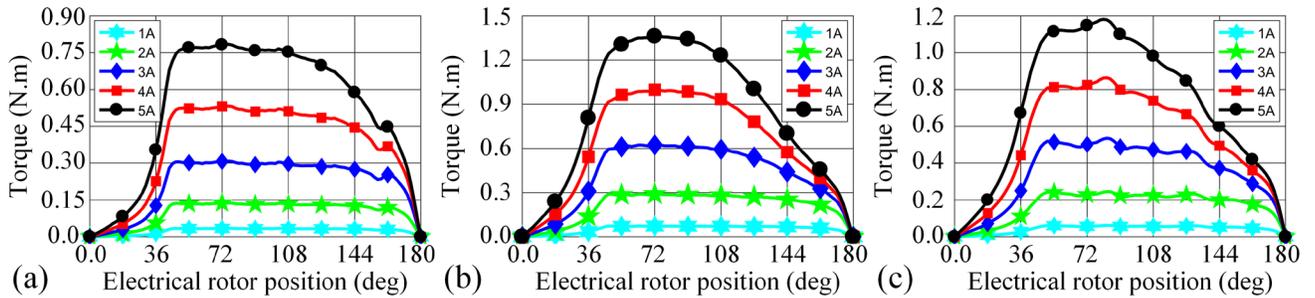

Fig. 4. Torque-angle characteristics.

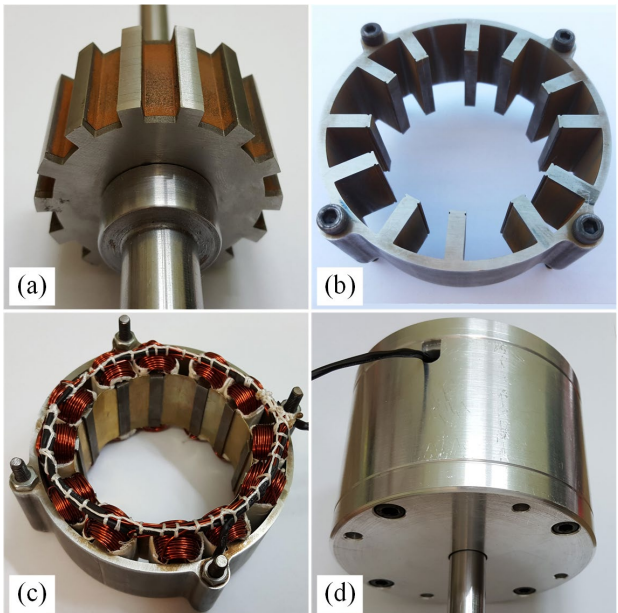

Fig. 5. Prototyped 12/14 SRM: (a) rotor, (b) stator core, (c) stator with windings, and (d) assembled motor.

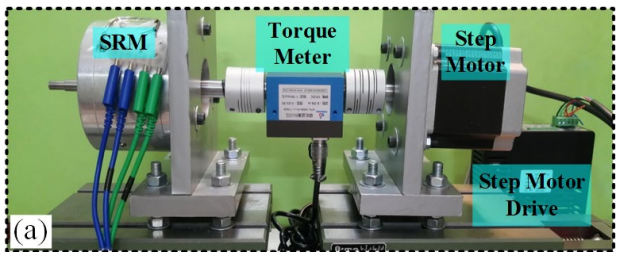

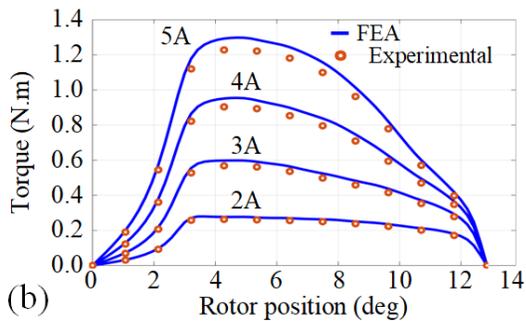

Fig. 6. Experimental study: (a) torque-angle measurement setup, and (b) the characteristic for the 12/14 SRM.

## III. RESULTS AND COMPARISONS

Fig. 3 illustrates magnetic flux density and flux lines for the three SRMs with an excitation of 5 A at phase A. It is seen that the highest magnetic flux density occurs in the C-core of excited windings. Fig. 4 shows the torque-angle characteristics of three SRMs at various current levels, along with mean and peak torques given in Table II. Table III compares the mean and peak torques of the selected 12/14 SRM with the conventional 12/8

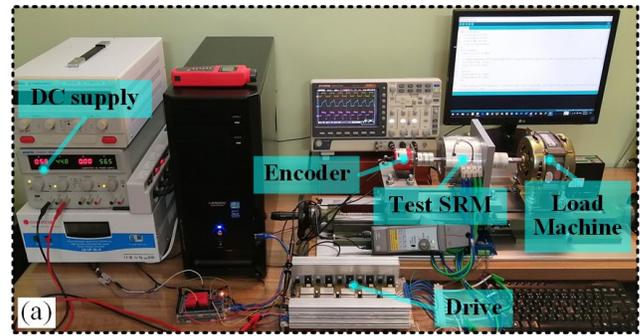

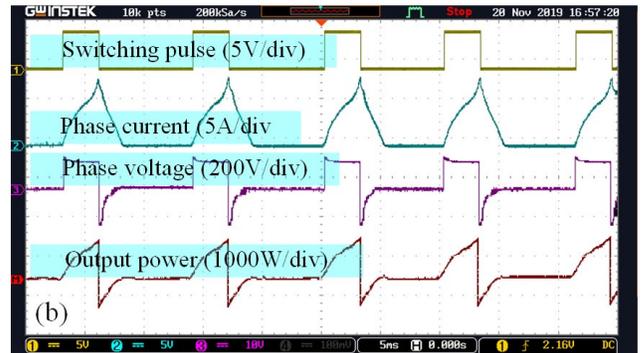

Fig. 7. Dynamic test: (a) experimental setup, (b) switching pulse, current, phase voltage, and input power waveforms at speed of 2000 rpm.

SRM, showing a 16.23% increase in mean torque at 5 A and confirming the superiority of the motor.

## I. PROTOTYPING AND EXPERIMENTAL RESULTS

The prototyped 12/14 SRM, shown in Fig. 5, is tested as illustrated in Fig. 6 (a), resulting in the torque-angle characteristic given in Fig. 6(b), which represents a good correlation with FEM results at different values of stator current. Fig. 7 shows the dynamic test setup for single-pulse control of the motor as well as the waveforms of the switching pulse, current, voltage, and input power of one phase at the speed of 2000 rpm. Table IV makes a comparison with the simulation results, revealing a great agreement.

TABLE III. TORQUE COMPARISONS

| Phase current (A) | Mean torque (N.m) | | Peak Torque (N.m) | | 12/14 SRM compared to 12/8 SRM | |
|---|---|---|---|---|---|---|
| | 12/14 SRM | 12/8 SRM | 12/14 SRM | 12/8 SRM | Mean torque increase (%) | Peak torque increase (%) |
| 1 | 0.051 | 0.038 | 0.070 | 0.077 | 25.49 | -10 |
| 2 | 0.195 | 0.148 | 0.278 | 0.311 | 24.10 | -11.87 |
| 3 | 0.389 | 0.309 | 0.597 | 0.648 | 20.56 | -11.91 |
| 4 | 0.597 | 0.489 | 0.954 | 1.032 | 18.09 | -8.17 |
| 5 | 0.807 | 0.676 | 1.297 | 1.435 | 16.23 | -10.63 |

TABLE IV. DYNAMIC TORQUE OF 12/14 SRM

| | Predicted | Measured |
|---|---|---|
| Speed, *rpm* | 2000 | 2012 |
| Torque, *N.m* | 0.783 | 0.776 |
| Output power, *W* | 163.99 | 163.50 |
| Total core loss *W* | 25.77 | 28.58 |
| Input power, *W* | 189.76 | 192.08 |
| Efficiency, % | 86.42 | 85.12 |

## II. Conclusion

The optimized topology for a three-phase C-core SRM is investigated to maximize the electrical loading capability and, thus, the torque density of the motor. The general formulations for any number of C-core per phase and any number of rotor teeth are discussed and analyzed. Three C-core SRMs with different combinations of rotor and stator teeth are optimized and compared. Using FEM, magnetic fields and obtained. Finally, the selected motor is prototyped, and both static and dynamic tests are carried out. When the phase excitation is turned off, the current returns to zero pretty fast, which is advantageous as it does not produce a negative torque. The results obtained from FEM and the experiment correlate well. Also, it confirms the superiority of the C-core topology over the conventional SRMs.